\documentclass[journal,onecolumn]{IEEEtran}
\usepackage{cite}
\usepackage{amsmath}
\usepackage{graphicx}
\usepackage{algorithm}
\usepackage{algpseudocode}
\usepackage{multirow}
\usepackage[table,xcdraw]{xcolor}
\usepackage{authblk}
\usepackage[bookmarks=false]{hyperref}
\usepackage{xcolor}

\usepackage{subcaption}

\algblock{Input}{EndInput}
\algnotext{EndInput}
\algblock{Output}{EndOutput}
\algnotext{EndOutput}

\newcommand*{\affaddr}[1]{#1} 
\newcommand*{\affmark}[1][*]{\textsuperscript{#1}}
\newcommand*{\email}[1]{\textit{#1}}

\begin{document}

\title{A Concern Analysis of FOMC Statements Comparing The Great Recession and The COVID-19 Pandemic\footnote{This paper is the pre-print of a paper to appear in the proceedings of the IEEE International Conference on BigData 2020 (Workshops) entitled: ``{\it A Concern Analysis of Federal Reserve Statements: The Great Recession vs. The COVID-19 Pandemic}.}}

\author{%
Luis Felipe Gutiérrez\affmark[1], Sima Siami-Namini\affmark[2], Neda Tavakoli\affmark[3], and Akbar Siami Namin\affmark[1]\\
\affaddr{\affmark[1, 2, 3]Department of Computer Science} \\
\affaddr{\affmark[1]Texas Tech University}
\affaddr{\affmark[2]Mississippi State University} 
\affaddr{\affmark[3]Georgia Institute of Technology} \\
\email{\{Luis.Gutierrez-Espinoza, akbar.namin\}@ttu.edu ; ss4625@msstate.edu ; neda.tavakoli@gatech.edu}\\
}

\IEEEoverridecommandlockouts
\IEEEpubid{\makebox[\columnwidth]{978-1-7281-6251-5/20/\$31.00~\copyright2020 IEEE \hfill} \hspace{\columnsep}\makebox[\columnwidth]{ }}

\maketitle

\begin{abstract}
It is important and informative to compare and contrast major economic crises in order to confront novel and unknown cases such as the COVID-19 pandemic. The 2006 Great Recession and then the 2019 pandemic have a lot to share in terms of unemployment rate,  consumption expenditures, and interest rates set by Federal Reserve. In addition to quantitative historical data, it is also interesting to compare the contents of Federal Reserve statements for the period of these two crises and find out whether Federal Reserve cares about similar concerns or there are some other issues that demand separate and unique monetary policies. This paper conducts an analysis to explore the Federal Reserve concerns as expressed in their statements for the period of 2005 to 2020. The concern analysis is performed using natural language processing (NLP)  algorithms and a trend analysis of concern is also presented. We observe that there are some similarities between the Federal Reserve statements issued during the Great Recession with those issued for the 2019 COVID-19 pandemic.


\end{abstract}

\begin{IEEEkeywords}
    Natural Language Processing, Concern Analysis, the COVID-19 Pandemic, the Great Recession. 
\end{IEEEkeywords}

\section{Introduction}

The COVID-19 has struck the economy of the world. People from many nations and countries, regardless of their economic standings, are suffering due to the catastrophic impact the coronavirus pandemic has had on their economy. Major giant industries and private sectors such as airlines, oil and gas, leisure facilities, and manufacturing have laid off majority of their employees or asked for furlough.  The consequences of such devastating economic situation need to be studied and proper remediation actions should be recommended to the authorities such as Federal Reserve or central banks. 

As a reasonable approach to deal with analyzing the economic impacts of COVID-19, it makes sense to compare this pandemic with previous and similar situations and take the lessons learned there and apply them here. As a comparable case with tons of learned lessons, the Great Recession that hit the financial markets during 2006 - 2009 can be studied and compared with the coronavirus pandemic. Then, the confrontation strategies and remediation the authorities employed during that period of time can be explored and adapted to minimize the impacts of COVID-19 on economy. 

For instance, we can look at the historical data captured by the Federal Reserve such as interest rates, Growth Domestic Product (GDP), inflation, and unemployment rate during the Great Recession and then explore their influence on the economy for the period after the Great Recession. Accordingly, the monetary and fiscal policies that were employed after the Great Recession can be adapted with some justification and adjustment for after COVID-19 era. In addition, to analyzing quantitative data collected by the Federal Reserve, it is also possible to do some other types of analysis using non-structured textual data such as the Federal Reserve statements prepared by the Federal Open Market Committee (FOMC). 

Concern and trend analysis is the application of natural language processing (NLP) algorithms on chronological and unstructured textual data. Through concern analysis it is possible to conduct complementary  analysis and capture the trends of financial or economical concerns over a given period of time. The building block of concern and then trend analysis is topic modeling (TM) where the candidate topics of a given text are captured automatically using NLP-based topic analysis. 

This paper compares Federal Reserve statements for the period between 2005 and 2020 with the goal of capturing the similarities of the Federal Reserve concerns between the Great Recession and the COVID-19 pandemic. To do so, we adapt Latent Drichlet Allocation (LDA) and further use two strategies such as Bag of Words (BoW) and the frequency-based approaches such as the term frequency -inverse document frequency ($tf-idf$) algorithms in detecting topics.  

The results of our study show that the consequences and economic impacts of COVID-19 are far deeper damaging than the Great Recession. During the period of COVID-19, the unemployment rate is rocket high (close to 15\%) and the interest rate is as low as zero. We present and compare the trend of concerns for these two cases (i.e., the Great Recession and COVID-19) and draw some conclusions. This paper makes the following key contributions:

\begin{itemize}
    \item[--] We capture the topic and concerns of Federal Reserve statements through NLP-based algorithms.
    \item[--] The paper compares and contrasts the economic impacts of the Great Recession and COVID-19 using concern analysis.
    \item[--] A quantitative comparison of the Great Recession and COVID-19 is presented using quantitative data.
\end{itemize}

The remainder of this paper is organized as follows: Section \ref{sec:relatedwork} reviews the related research work in this line of research. In Section \ref{sec:background} the technical background of the NLP-based algorithms utilized in this work are presented. Section \ref{sec:setup} the experimental setup and the data collection procedure are explained. We present the results of our study in Section \ref{sec:results}. Section \ref{sec:notablefinding} highlights our economic findings and notable observations in this study. The economic impacts of the Great Recession and the pandemic are compared in Section \ref{sec:Impacts}. Section \ref{sec:conclusion} concludes the paper and highlights the future research work.

\section{Related work}
\label{sec:relatedwork}

FOMC regularly holds eight scheduled meetings during the year to assess the economy indicators and makes key decisions about interest rates and the growth of the U.S. money supply in response to the severe crises such as the Great Recession occurred in 2006 and nowadays, the COVID-19 pandemic outbreak. The long-run goals are to achieve maximum employment, stable inflation at two percents, stimulate economic growth and stable financial markets. The Federal Reserve  releases a statement after each FOMC meetings to set expectations about  monetary policy. In fact, the Federal Reserve is responsible for setting the interest rate, which influences portfolio choice and asset prices. For example, the Federal Reserve had decreased interest rates and adopted unconventional monetary policy in response to the economic downturn of the Great Recession. Now, the Federal Reserve has decreased interest rates in response to the COVID-19 pandemic outbreak. The open question is how  the FOMC statements have been changed over time? How effective are the statements in making expectations? How can we predict the Federal Reserve's decisions? 

To address these questions, several methods are used to analyze the Federal Reserve statements and expectations. Among them, NLP-based algorithms and text analysis are recently used to compare the discussed words and topics in Federal Reserve statements. The tf-idf weighting method can provide higher weights to the terms that appear to carry more information from the Federal Reserve statements. In fact, keyword extraction such as topic detection and tracking is the most fundamental tasks in the field of text mining and NLP-based algorithms \cite{Kamalrudin}.

Several Topic Modeling (TM) methods are used to extract topics from short- and long- texts \cite{Xie, Cheng} which include Probabilistic Latent Semantic Analysis (PLSA) \cite{Hofmann}, Latent Semantic Analysis (LSA) \cite{Deerwester} and Latent Dirichlet allocation (LDA) by \cite{Blei2}. However, there are some issues in using TM methods such as data sparsity, spelling and grammatical errors, noisy words, and unstructured data, which need to be removed first. For example, Biterm Topic Model (BTM) is an advanced TM method that uses word correlations \cite{Yan}. The focus of this paper is LDA analysis and this section reports some related works in this field. LDA analysis is recently used by researchers at central banks to identify topic priorities in the bank statements. 

Ramachandran and DeRose \cite{Ramachandran} created a semantic space of the cumulative perspective of the FOMC meetings in 2017 by using LSA method. They utilized the cosine similarity as a measure of finding correlation between speech and minutes in a high dimensional space and identified the similarity between  policymakers and the committee consensus. The results showed that three policymakers including Kaplan  with 0.67, Yellen with 0.65 and Evans with 0.61 have the highest correlation on average, respectively. 

Edison and Marquez \cite{Edison2} analyzed the FOMC transcripts for the period of 2003-2012 (including 45,346 passages) by using LDA statistical models and machine learning algorithms. They found the evolution of eight different topics including ‘forecasting’, ‘banking system’, ‘economic modeling’, ‘voting decision’, ‘statement language’, ‘economic activity’, ‘risks’, and ‘communication’. The results showed that ‘economic modelling’ was dominant during the Great Recession and financial crisis, with an increase in the ‘banking system’ in the following years, and ‘communication’ in the recent years. However, the evolution of some economic terminology such as Taylor rule \cite{Taylor93} or change in the general tone in the statements need to be investigated. 

Albalawi et al.\ \cite{Albalawi} compared different TM methods for short-text data analysis by using two textual datasets including the 20-newsgroup data ($20,000$ documents) and short-text data from the Facebook website (20 text conversations). The results showed that all TM methods including LSA, LDA, Non-negative Matrix Factorization (NMF), Principal Component Analysis (PCA), and RP are similar in transferring text into term, document frequency matrices, using the tf-idf method, producing topic content weights for each document, but LDA and NMF methods provided the best results with diverse ranges and meanings. 

Du et al. \cite{8754045} applied LDA to analyze people concerns during the 2018 California Wildfire using two sources of texts: news websites and Twitter. In addition, the retrieved concerns were analyzed in terms of importance and how they evolved over time. The results showed that even though the main focus was the wildfire, the concerns present high variations, as the texts tend to focus on different aspects of the event.

\section{Technical Background}
\label{sec:background}

\subsection{Bag-of-Words and tf-idf}

The purpose of the tf-idf technique in a document-term frequency matrix is to weigh terms in such a way that rare terms over documents hold higher values, and common ones have lower values \cite{turney2010frequency}.

Let $A$ be a $d \times t$ document-term frequency matrix, known as the BOW matrix, where $d$ is the number of documents in the dataset and $t$ is the number of unique terms. Each entry $A_{i,j}$ contains the frequency of the term $t_j$ in the document $d_i$. The tf-idf processing generates a new matrix $T$ where each entry is computed as follows \cite{ramos2003using}:

 \vspace{-0.1in}
\[
    T_{i,j} = A_{i,j} \cdot idf(j),
\]

The factor $idf(j)$ is calculated as 

 \vspace{-0.1in}
\[
    idf(j) = \log \left ( \frac{d}{f_j} \right ),
\]

\noindent $f_j$ is the number of documents in which the term $t_j$ is present.

In order to allow more interpretability, we excluded from the BOW matrix all words that appeared in more than 50\% of our dataset.

\subsection{Latent Dirichlet Allocation (LDA)}
\label{sec:compared}

LDA is a technique used for topic modelling over a collection of documents \cite{blei2003latent}. In this generative probabilistic model, each document in a corpus is modeled as a mixture of latent topics. At the same time, each topic is modeled as a distribution over words of the dictionary. 

\begin{figure}[h]
  \centering
  \includegraphics[width=0.9\linewidth]{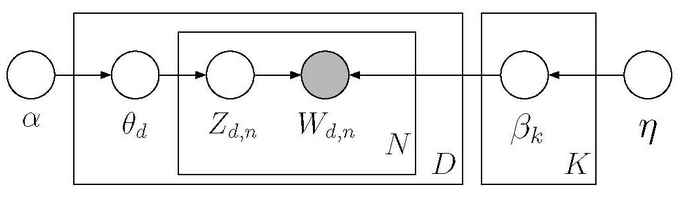}
  \caption{Graphical model of LDA (adapted from \cite{blei2003latent}).}
  \label{fig:lda-model}
   \vspace{-0.2in}
\end{figure}

Figure \ref{fig:lda-model} depicts the graphical representation of the LDA model, the outer box represents the generative process for each one of the M documents, while the inner box represents the generation of topics given the word probabilities $\beta$. This model considers the following parameters:

\begin{itemize}
    \item[--] $\alpha$: a $k$-dimensional vector, with $\alpha_i > 0$ for $i = {1...k}$. It is used to sample the $\Theta$ parameter for each document in the corpus.
    \item[--] $\beta_d$: a $k \times V$ matrix representing word probabilities over topics, where $V$ is the size of the dictionary.
    \item[--] $\eta$: Likewise $\alpha$, it is used to determine the prior probabilities of the topic-words probabilities.
    \item[--] $\Theta_d$: a $k$-dimensional Dirichlet variable, $\Theta \sim Dir(\alpha)$, also known as topic mixture, this is specific to document $d$.
    \item[--] $N$: Collection of words.
    \item[--] $M$: Collection of documents.
    \item[--] $K$: Collection of topics.
    
\end{itemize}

According to the frequency of sampling, the parameters of this model can be categorized into three different levels: 1) Corpus level, 2) Document level, and 3) Word-level \cite{blei2003latent}.

\subsection{Topic Coherence Measures}

In our study, we use the LDA topic coherence measure framework  \cite{roder2015exploring}. The framework consists of 

\begin{enumerate}
    \item {\it Segmentation.} An original set of words $W$ (a topic in this analysis) is divided into several subsets (e.g., pair of words). Each subset will be compared to each other in subsequent steps. The resulting set of segmentations is denoted by $S$.
    \item {\it Probability Estimation.} This step defines the method that will be used to estimate the word probabilities under the source data, such as probabilities of a single word, or the joint probability given a pair of words. The set of all probabilities is $\mathcal{P}$.
    \item {\it Confirmation Measure.} Given the corresponding $S$, or a pair of members of $S$, and the probablities $\mathcal{P}$, a confirmation measure is calculated according to the support between the members of $S$. The set of confirmation measures is denoted as $\mathcal{M}$.
    \item {\it Aggregation.} In order to compute the final coherence score, the confirmation measures must be aggregated in the some way. The set of aggregations is denoted as $\Sigma$.
\end{enumerate}

Using all the components of the framework, the set of all possible coherence measures $\mathcal{C}$ is given by $S \times \mathcal{P} \times \mathcal{M} \times \Sigma$. Several coherence measures can be calculated by varying the choice for each component of the framework. A discussion of different coherence measures can be found in \cite{roder2015exploring}. 

In this work, we used the $C_v$ coherence measure, hereafter referred to as {\it coherence score}, which has reported the highest average score in \cite{roder2015exploring}. $C_v$ ranges from 0 to 1, lower values of $C_v$ suggest a set of topics that is hard to interpret, whereas values close to 1 allow an easier human interpretation of the topics. Furthermore, we use the coherence score to discriminate between models and perform model selection.

\section{Experimental Setup}
\label{sec:setup}

\subsection{Libraries}

We developed Python 3 scripts in order to execute our experiments. We used the Gensim library \cite{rehurek_lrec} to implement the BOW, tf-idf, LDA models, and the coherence score. For named entity recognition in the data preprocessing step, we utilized the spaCy library \cite{spacy2}. For visualization of the LDA topics, we utilized the pyLDAvis library \cite{sievert2014ldavis}. For stemming, we used the Porter stemmer algorithm provided by the NLTK toolkit \cite{loper2002nltk}.

\subsection{Data Collection}

Our dataset consists of 127 uninterrupted FOMC's postmeeting statements. The first statement was issued on February 2005; whereas, the last one was issued on July 2020. We developed Python scripts for retrieving the documents directly from the website of the Federal Reserve \footnote{https://www.federalreserve.gov/newsevents.htm}. 

\subsection{Data Preprocessing and Normalization}

We performed widely adopted data cleansing and normalization techniques in natural language processing. We performed the following steps:

\begin{itemize}
    \item[--] Remove any special characters resulting from the web scrapping.
    \item[--] Using spaCy, apply named entity recognition to each statement and remove any referred name for both people and cities.
    \item[--] Remove punctuation characters.
    \item[--] Transform the statements to lower cases.
    \item[--] Remove frequent words determined beforehand (e.g., board, approve, governor, etc).
    \item[--] Using the Porter stemmer algorithm provided by the NLTK library, stem each token of the statements.
\end{itemize}

\subsection{Hyperparameter Tuning}

We tuned the following hyperparameters in the LDA model: number of topics, $\alpha$, and $\eta$. When dealing with the interpretability of LDA, the number of topics is the most important hyperparameter; nevertheless, as this problem is unsupervised, there is not a ground truth to which we can compare. 

In order to explore the performance of the model, we tried several different configurations of hyperparameters and reported their coherence scores. We performed a grid search for the values of the number of topics, $\alpha$, and $\eta$ according to the ranges:

\begin{itemize}
    \item[--] Number of topics: 3 to 10, increments of 1.
    \item[--] $\alpha$: 0.05 to 1.55, increments of 0.1.
    \item[--] $\eta$: 0.05 to 1.55, increments of 0.1.
\end{itemize}

However, since the Cartesian product between the three ranges has 1800 elements, we randomly chose 100 sets of hyperparameters so we could fit the LDA models in a more limited time.

\subsection{Methodology Flowchart}

\begin{figure}[h]
  \centering
  \includegraphics[width=\linewidth]{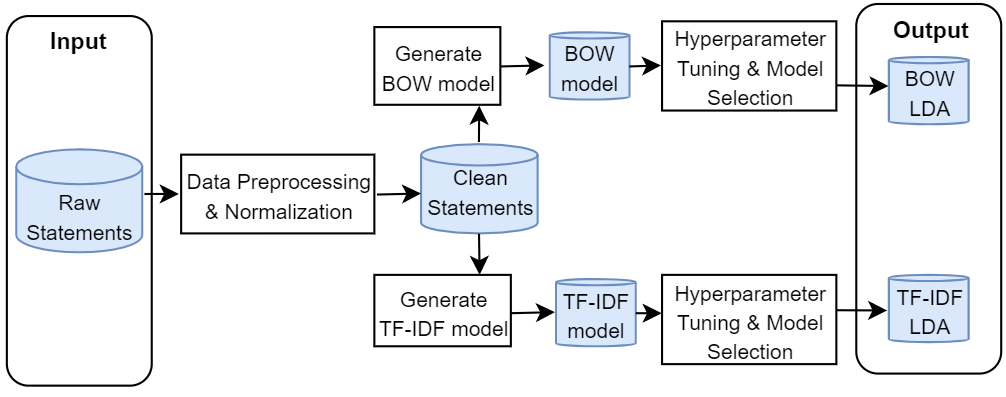}
  \caption{Flowchart of the methodology applied to our work.}
  \label{fig:method-flowchart}
   \vspace{-0.2in}
\end{figure}

Figure \ref{fig:method-flowchart} shows the flowchart for the experiments in this work. The input of the pipeline is the raw statements, while the final output is the LDA models for both BOW and tf-idf representations of documents. The final LDA models are obtained after the hyperparameter optimization and LDA model selection using the highest coherence scores as the criteria for selection, as was explained in the previous section.

\section{Results}
\label{sec:results}

\subsection{Hyperparameters Selection}

Tables \ref{tab:coherence-bow} and \ref{tab:coherence-tfidf} show the top 10 coherence scores using LDA with a BOW model and a tf-idf model, respectively, alongside with the set of hyperparameters. Table \ref{tab:coherence-bow} shows that for the BOW model, the highest coherence score is achieved using 5 topics, with $\alpha = 0.55$ and $\eta = 0.45$. On the other hand, Table \ref{tab:coherence-tfidf} shows that the highest coherence score using the tf-idf model is reached with 10 topics, with $\alpha = 0.55$ and $\eta = 1.15$. It is worth noting that while high coherence scores are achieved with the number of topics ranging between 5 and 8 in the case of the BOW model, the tf-idf model consistently reports the highest scores using 9 or 10 topics. In addition, scores for tf-idf models are higher than those of BOW models, where the highest score for tf-idf is 23\% higher than the BOW one.

\begin{table}[h]
    \centering
    \begin{tabular}{|c|r|r|c|}
    \hline
     {\bf N Topics} &  {\bf $\alpha$} & {\bf $\eta$} &  {\bf Coherence BOW} \\
     \hline
            5 &   0.55 &  0.45 &       0.550850 \\
            5 &   0.65 &  1.25 &       0.546334 \\
            8 &   1.45 &  0.35 &       0.544440 \\
            7 &   1.05 &  0.25 &       0.543520 \\
            6 &   1.05 &  1.15 &       0.542821 \\
            9 &   1.25 &  0.05 &       0.541682 \\
            6 &   0.65 &  1.25 &       0.539881 \\
            5 &   0.35 &  1.25 &       0.539420 \\
            7 &   0.25 &  0.35 &       0.538040 \\
            5 &   0.95 &  0.85 &       0.537979 \\
    \hline
    \end{tabular}

    \caption{Top 10 Coherence scores: LDA using BOW model.}
    \label{tab:coherence-bow}
     \vspace{-0.1in}
\end{table}

\begin{table}[h]
    \centering
    \begin{tabular}{|c|r|r|c|}
    \hline
         {\bf N Topics} &  {\bf $\alpha$} & {\bf $\eta$} &  {\bf Coherence tf-idf} \\
    \hline
           10 &   0.55 &  1.15 &         0.686169 \\
           10 &   0.65 &  0.75 &         0.685247 \\
            9 &   0.55 &  1.15 &         0.674652 \\
            9 &   0.75 &  0.65 &         0.663492 \\
            6 &   0.75 &  1.05 &         0.648774 \\
            6 &   0.65 &  1.25 &         0.639563 \\
            5 &   0.75 &  1.45 &         0.631015 \\
            6 &   1.05 &  0.45 &         0.623710 \\
            4 &   0.65 &  1.05 &         0.616420 \\
            3 &   0.35 &  1.45 &         0.615097 \\
    \hline
    \end{tabular}

    \caption{Top 10 Coherence scores:LDA using tf-idf model.}
    \label{tab:coherence-tfidf}
     \vspace{-0.1in}
\end{table}

Onwards, our analysis was conducted using the LDA models with the hyperparameters reporting the highest coherence score for both BOW and tf-idf models.

\begin{figure*}[!ht]
      \centering
    \begin{subfigure}{.19\textwidth}
      \centering
      \includegraphics[width=\textwidth]{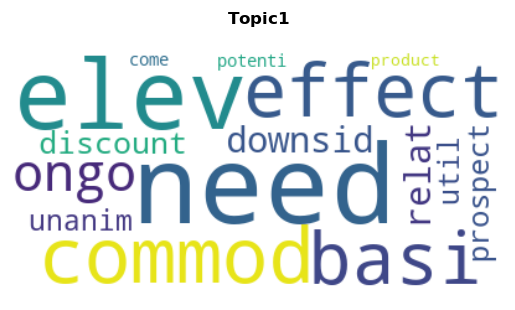}
      \caption{Financial Market (Topic 1)}
      \label{fig:wc-topic1-bow-7t}
    \end{subfigure}%
    \enspace
    \begin{subfigure}{.19\textwidth}
      \centering
      \includegraphics[width=\textwidth]{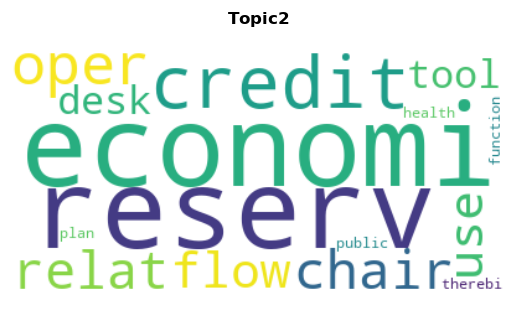}
      \caption{Healthcare Plan (Topic 2)}
      \label{fig:wc-topic2-bow-7t}
    \end{subfigure}%
        \enspace
    \begin{subfigure}{.19\textwidth}
      \centering
        \includegraphics[width=\textwidth]{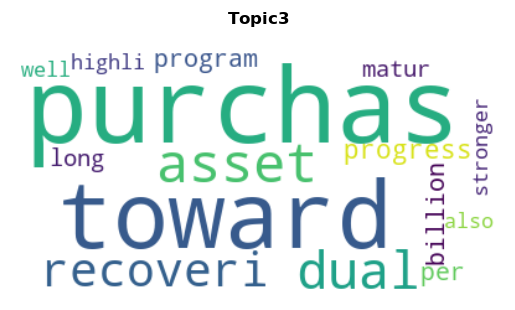}
        \caption{Asset Purchase and Recovery (Topic 3)}
        \label{fig:wc-topic3-bow-7t}
    \end{subfigure}%
        \enspace
    \begin{subfigure}{.19\textwidth}
      \centering
        \includegraphics[width=\textwidth]{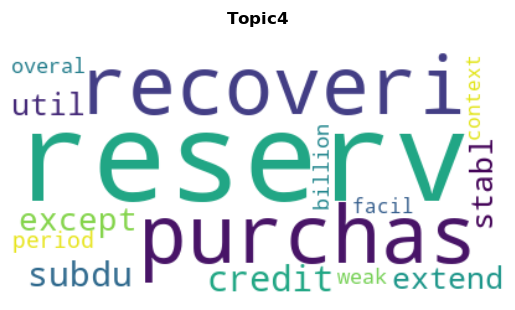}
        \caption{Credit (Topic 4)}
        \label{fig:wc-topic4-bow-7t}
    \end{subfigure}%
        \enspace
    \begin{subfigure}{.19\textwidth}
      \centering
        \includegraphics[width=\textwidth]{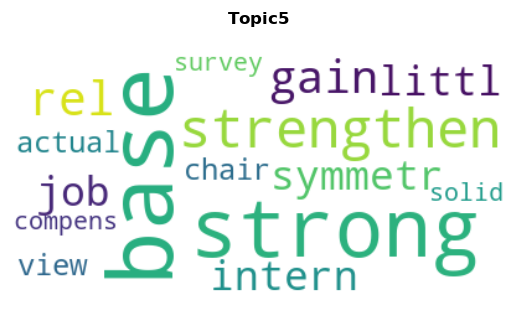}
        \caption{Labor Market (Topic 5)}
        \label{fig:wc-topic5-bow-7t}
    \end{subfigure}%
    \caption{Wordclouds for the LDA topics. BOW Model. N topics = 5.}
  \label{fig:wc-bow-7t}
   \vspace{-0.2in}
\end{figure*}

\begin{figure*}[!ht]
      \centering
    \begin{subfigure}{.19\textwidth}
      \centering
      \includegraphics[width=\textwidth]{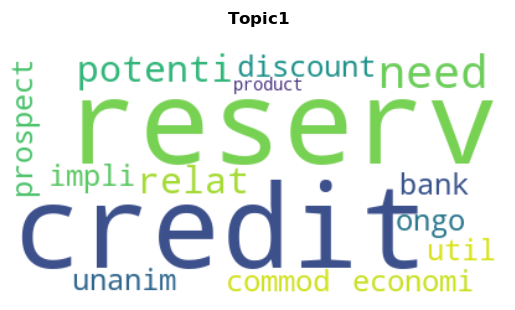}
      \caption{General Monetary Policy (Topic 1)}
      \label{fig:wc-topic1-tfidf-10t}
    \end{subfigure}%
            \enspace
    \begin{subfigure}{.19\textwidth}
      \centering
      \includegraphics[width=\textwidth]{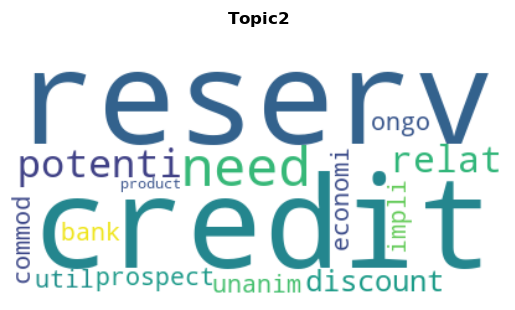}
      \caption{General Monetary Policy (Topic 2)}
      \label{fig:wc-topic2-tfidf-10t}
    \end{subfigure}%
            \enspace
    \begin{subfigure}{.19\textwidth}
      \centering
      \includegraphics[width=\textwidth]{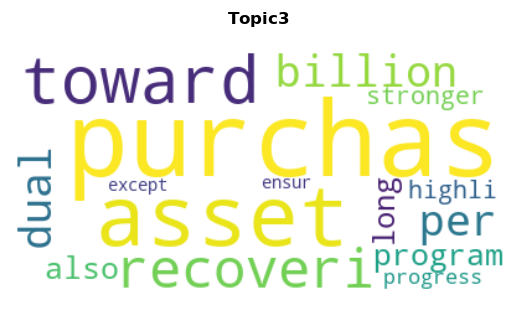}
      \caption{Asset Purchase and Recovery (Topic 3)}
      \label{fig:wc-topic3-tfidf-10t}
    \end{subfigure}%
            \enspace
    \begin{subfigure}{.19\textwidth}
      \centering
      \includegraphics[width=\textwidth]{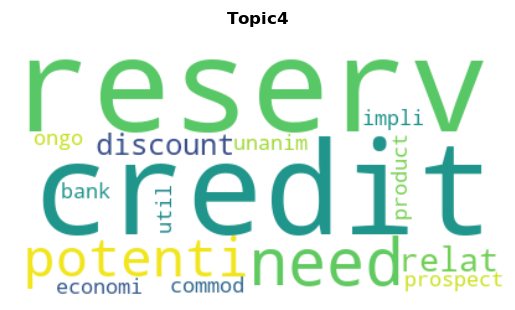}
      \caption{General Monetary Policy (Topic 4)}
      \label{fig:wc-topic4-tfidf-10t}
    \end{subfigure}%
            \enspace
    \begin{subfigure}{.19\textwidth}
      \centering
        \includegraphics[width=\textwidth]{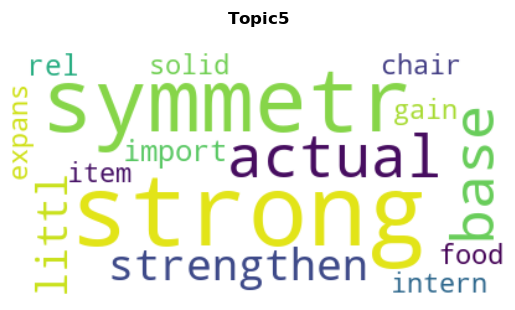}
        \caption{Economic Growth (Topic 5)}
        \label{fig:wc-topic5-tfidf-10t}
    \end{subfigure}%
    
    \begin{subfigure}{.19\textwidth}
      \centering
        \includegraphics[width=\textwidth]{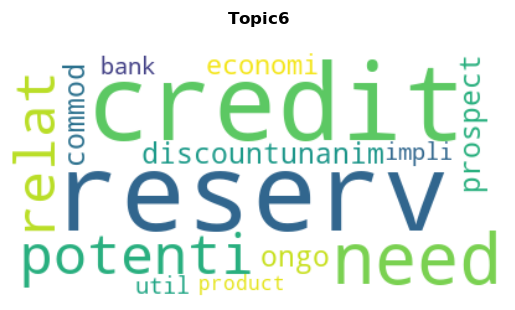}
        \caption{General Monetary Policy (Topic 6)}
        \label{fig:wc-topic6-tfidf-10t}
    \end{subfigure}%
            \enspace
    \begin{subfigure}{.19\textwidth}
      \centering
        \includegraphics[width=\textwidth]{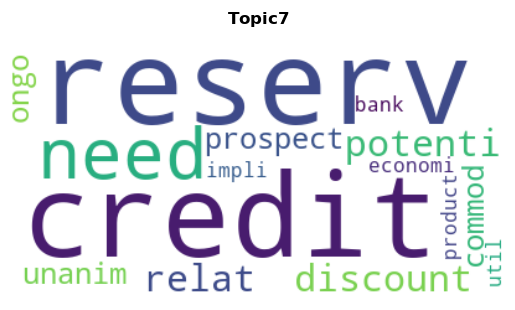}
        \caption{General Monetary Policy (Topic 7)}
        \label{fig:wc-topic7-tfidf-10t}
    \end{subfigure}%
            \enspace
    \begin{subfigure}{.19\textwidth}
      \centering
        \includegraphics[width=\textwidth]{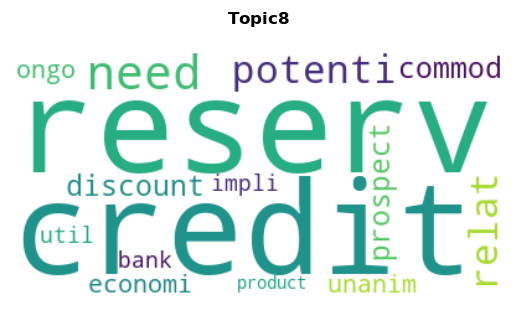}
        \caption{General Monetary Policy (Topic 8)}
        \label{fig:wc-topic8-tfidf-10t}
    \end{subfigure}%
            \enspace
    \begin{subfigure}{.19\textwidth}
      \centering
        \includegraphics[width=\textwidth]{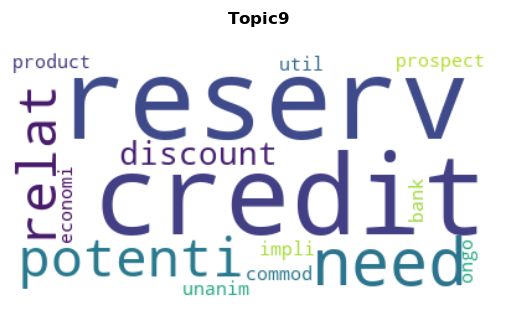}
        \caption{General Monetary Policy (Topic 9)}
        \label{fig:wc-topic9-tfidf-10t}
    \end{subfigure}%
            \enspace
    \begin{subfigure}{.19\textwidth}
      \centering
        \includegraphics[width=\textwidth]{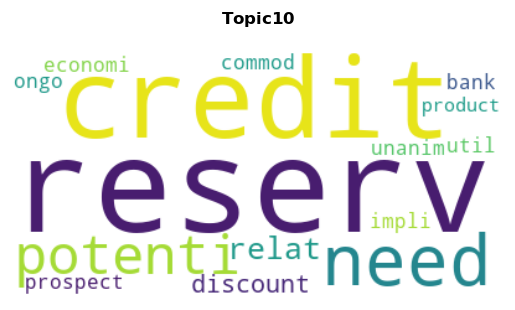}
        \caption{General Monetary Policy (Topic 10)}
        \label{fig:wc-topic10-tfidf-10t}
    \end{subfigure}
    
    \caption{Wordclouds for the LDA topics. tf-idf Model. N topics = 10.}
  \label{fig:wc-tfidf-10t}
   \vspace{-0.1in}
\end{figure*}

\subsection{Visualization of Topics}

\subsubsection{Wordcloud} Figures \ref{fig:wc-bow-7t} and \ref{fig:wc-tfidf-10t} show the wordclouds for the topics obtained using LDA with the BOW model and tf-idf model, respectively. Figure \ref{fig:wc-bow-7t} depicts the 15 most important words appeared in each topic using BOW model. The topics and their classifications are clearly different with some minor overlapping terms. 

Unlike different topics produced by the BOW model, the wordclouds representing the 10 topics created for tf-idf model are very similar. In other words, 8 out of 10 topics are somehow similar and they are labeled as ``{\it General Monetary Policy}''. Only two topics (Topics 3 and 5) are different than the other topics. This may indicate that the computation performed on measuring the coherence score based on tf-idf might need some revision in order to exclude general monetary policy terms from FOMC statements.

\subsubsection{Multidimensional Scaling (MDS)} 
\begin{figure*}[ht]
      \centering
    \begin{subfigure}{.4\textwidth}
      \centering
      \includegraphics[width=\linewidth]{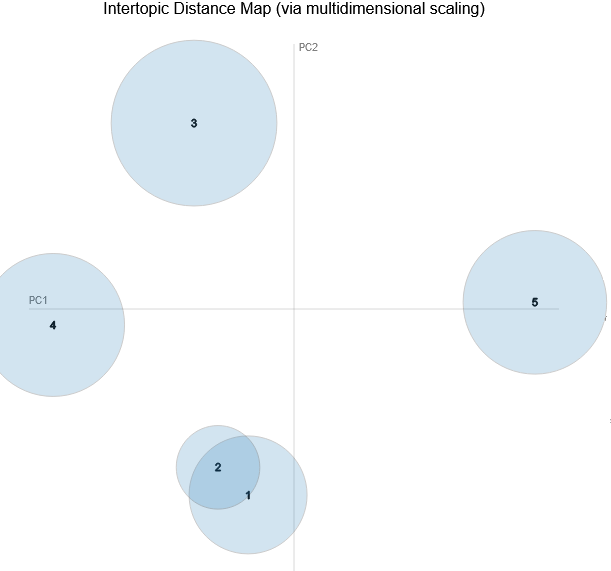}
      \caption{BOW Model. N topics = 5.}
      \label{fig:mds-topics-bow-7t}
    \end{subfigure}%
    \qquad
    \begin{subfigure}{.4\textwidth}
      \centering
      \includegraphics[width=\linewidth]{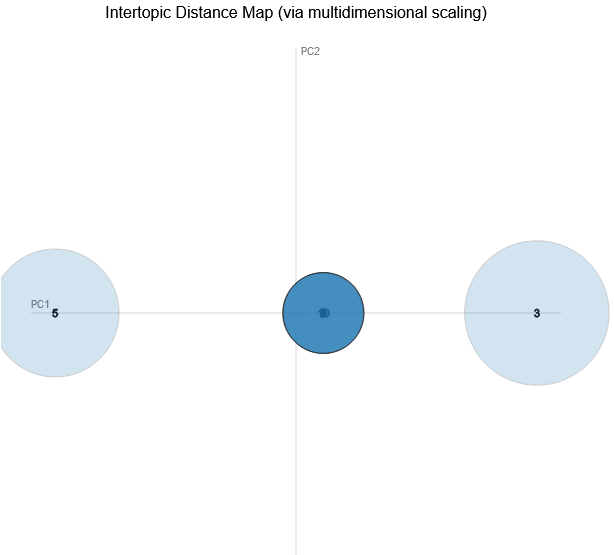}
      \caption{tf-idf Model. N topics = 10.}
      \label{fig:mds-topics-tfidf-10t}
    \end{subfigure}
  \caption{MDS representation of LDA topics.}
  \label{fig:mds-topics}
   \vspace{-0.2in}
\end{figure*}

Figure \ref{fig:mds-topics} shows the topic embeddings provided by pyLDAvis. These embeddings are generated using Multidimensional Scaling (MDS) on the high-dimensional topic vectors. The size of the blobs for each topic encodes its marginal topic distribution (i.e., how often and dominant the topic appears in the topic document topic mixtures, bigger blobs denote more dominant topics). 


Figure \ref{fig:mds-topics-bow-7t} shows that Topics 1 and 2 for BOW have some overlap, sharing similarity at some extent. In contrast, Topics 3, 4, and 5 are well separated in the plot, suggesting different concepts among them. The figure supports our observations from the wordclouds (Figure \ref{fig:wc-bow-7t}) where we reported BOW has created a heterogeneous set of topics.

On the other hand, Figure \ref{fig:mds-topics-tfidf-10t} shows that only topics 3 and 5 are significantly different from the others. This fact agrees with the wordclouds for Topics 1, 2, 4, 6, 7, 8, 9, 10 of the tf-idf model in the wordclous Figure \ref{fig:wc-tfidf-10t}, which are almost identical. However, although the Topics 1, 2, 4, 6, 7, 8, 9, 10 are overlapping, their dominance in the documents is not negligible.

\section{Notable Economic Findings: Topic Dominance in Statements Over Time}
\label{sec:notablefinding}

Figure \ref{fig:sp-bow-7t} and Figure \ref{fig:sp-tfidf-10t} show the stack plots of the topic dominance for the documents over time for BOW model and tf-idf model, respectively.

\subsection{BOW: Topic Dominance and Topic Labeling}
Figure \ref{fig:sp-bow-7t} shows a clear periodical topic dominance captured by the BOW model for the period of 2005 - 2020.  According to the BOW model, there are five distinct and clear trends during this period, as follows: 

\begin{itemize}
    \item[--] {\it 2005 - 2008.} In this period, Topic 1 is identified as the dominant topic. Topic 1 includes terms such as commodity, downside, discount, and prospect. During this period, the U.S. economy has been followed by a severe economic downturn, collapse of the housing market, the subprime mortgage crisis and bank failures. The immediate consequence was Great Recession. The Federal Reserve responded to the severe recession by cutting interest rate near zero in several steps to recover economy and preserve price stability. Therefore, it makes sense to label Topic 1 as ``{\it Financial Market.}''
    
    \item[--] {\it 2009 - 20011.} Topic 4 is the dominant topic here. The key terms appeared in Topic 4 include purchase, recovery, credit, extend, and weak. During this period, the American Recovery and Reinvestment Act of 2009 was signed. This Act was a stimulus package, which aimed to save existing jobs and create new ones, and invest in infrastructure, education, health, and renewable energy. During these years, the Federal Reserve purchase Treasury securities and agency mortgage-backed securities (MBS) to support the flow of credit to household and businesses. Then it makes sense to label Topic 4 as ``{\it Credit.}” 

    \item[--] {\it 2012 - 2015.} Topic 3 is the dominant topic in this period of time. The terms appeared in Topic 3 include purchase, asset, recovery, progress, well, and stronger.  During these years, the U.S. economy was recovering. The FOMC continued the large-scale asset purchase to extend the average maturity of Treasury securities. We labeled Topic 3 as ``{\it Asset Purchase and Recovery.}''
    
    \item[--] {\it 2016 - 2020.} Topic 5 is the dominant topic in this period. Topic 5 includes terms such as base, strong, strength, gain, job, and solid. It is when President Donald Trump took over the oval office. During this period of time the economy was rapidly growing, and unemployment rate decreased. We named Topic 5 as ``{\it Labor Market.}''
    
    \item[--] {\it 2020 - present.} Topic 2 is the dominant topic here. Topic 2 includes terms such as credit, flow, public, plan, and health. This is the period that the COVID-19 pandemic started to negatively impact the world’s economy. Accordingly, Topic 2 is labeled as ``{\it Healthcare Plan.}''

\end{itemize}


\subsection{tf-idf: Topic Dominance and Topic Labeling}

\begin{figure*}[t]
  \centering
  \includegraphics[width=\linewidth]{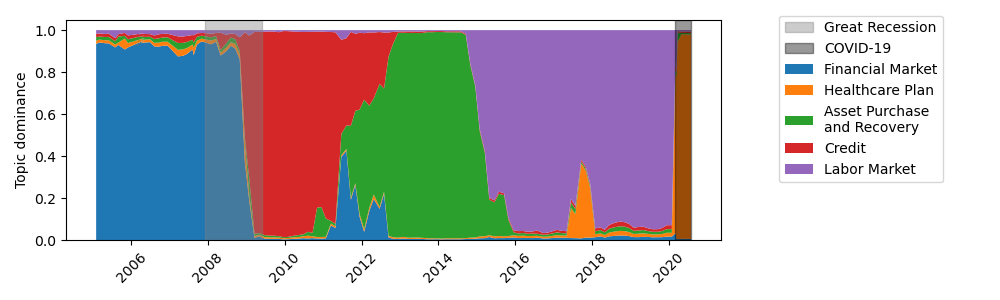}
  \caption{Stack plot of topic dominance over time. BOW Model. N of topics = 5.}
  \label{fig:sp-bow-7t}
   \vspace{-0.2in}
\end{figure*}

\begin{figure*}[t]
  \centering
  \includegraphics[width=\linewidth]{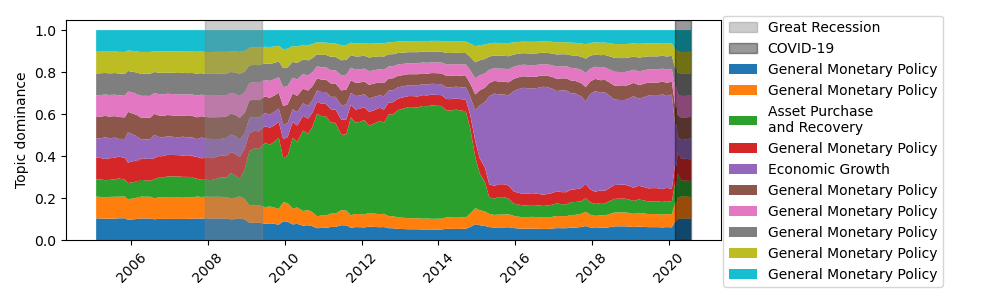}
  \caption{Stack plot of topic dominance over time. tf-idf Model. N of topics = 10.}
  \label{fig:sp-tfidf-10t}
   \vspace{-0.2in}
\end{figure*}

Unlike  the informative and meaningful patterns demonstrated by BOW in Figure \ref{fig:sp-bow-7t}, Figure \ref{fig:sp-tfidf-10t} shows a less informative pattern for demonstrating the trends during the period of 2005 and 2020. by the end of the Great Recession, the mixture gets clearly dominant towards Topic 3 until 2015, when it turns dominant towards Topic 5. Clearly, there are only two dominating topics:

\begin{itemize}
    \item[--] 2009 - 2015. The Great Recession occurred during this period and the Federal Reserve adopted zero-bound interest rate policy and then selected several rounds of quantitative easing (QE) policy to respond to the severe Great Recession. Topic 3 is the dominating topic in this period. It includes terms such as purchase, asset, recovery, stronger, program, and ensure. As stated above, this period consists of starting the Great Recession followed by economic recovery. Accordingly, Topic 3 is called ``{\it Asset Purchase and Recovery.}''
    
    \item[--] 2016 - 2020. This is the period when President Donald Trump took the office. Topic 5 is dominating this period and this topic consists terms such as strong, strength, solid, gain, food, and expansion. During this period the economy had tremendous growth and the Federal Reserve increased the short-term interest rate. Then, we labeled Topic 5 as ``{\it Economic Growth.}''

\end{itemize}

By comparing Figures \ref{fig:sp-bow-7t} and \ref{fig:sp-tfidf-10t} and the above economical explanation, 
we observe that the BOW model has greatly captured the distinct topics and thus has provided a better topic modeling than tf-idf. The model produced by tf-idf is still useful since it has also captured two major events 1) the Great Recession, and 2) the start of the President Donald Trump period. However, the tf-idf model fails in capturing anything interesting for the period of COVID-19 pandemic. All topics have shown up with equal domination magnitudes making hard to infer whether any concerns or trends are captured.

\section{Economic Impacts: The Great Recession vs. The COVID-19 Pandemic}
\label{sec:Impacts}

As mentioned earlier, the Federal Reserve meets eight times a year to determine the direction of the monetary policy. The Federal Reserve creates two sets of text data including 1) the statement, which is released at the moment of the target rate decision, and 2) the minutes which is released with a three-week lag. In this research, we detect the evolution of the different topics discussed in the statements for the period 2005-2020 covering the Great Recession and the COVID-19 pandemic. All statements are available in the Federal Reserve Website. 

\begin{figure}[!h]
      \centering
    \begin{subfigure}{0.8\linewidth}
      \centering
  \includegraphics[width=\linewidth]{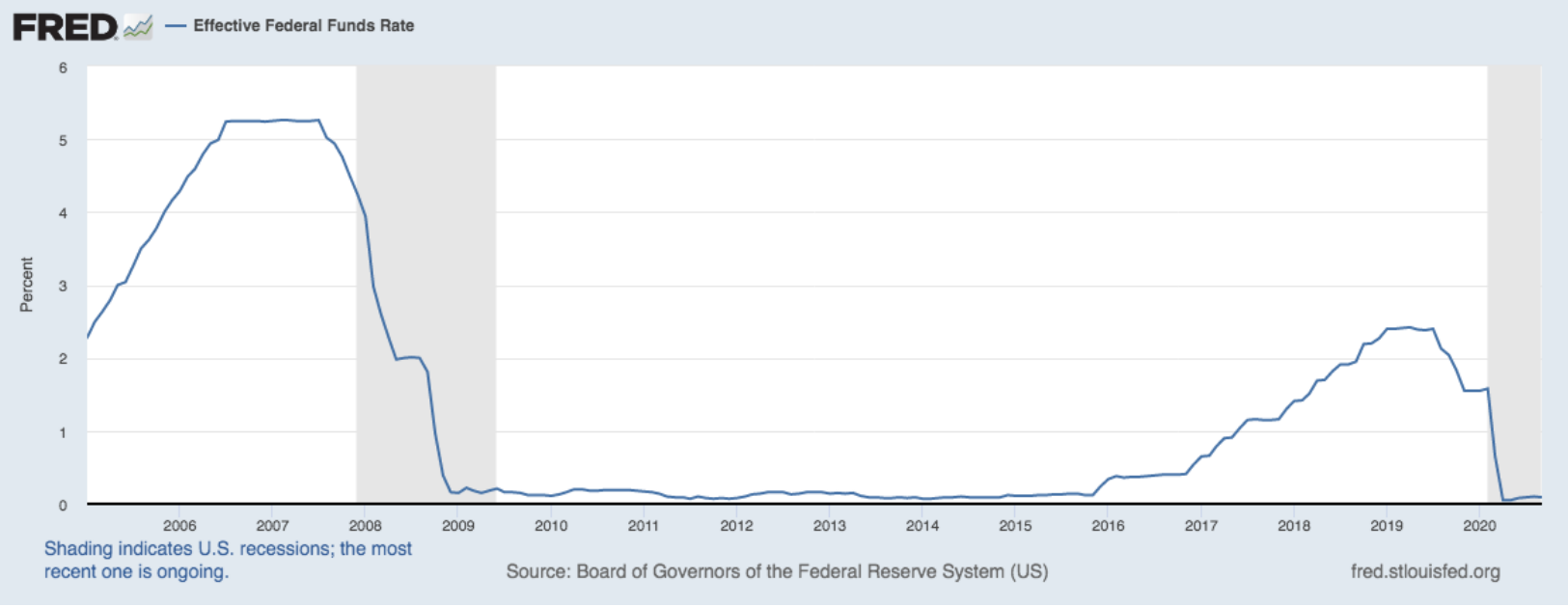}
  \caption{Monthly data of U.S. interest rate.}
  \label{fig:interestrate}
    \end{subfigure}%
    
    \begin{subfigure}{0.8\linewidth}
      \centering
  \includegraphics[width=\linewidth]{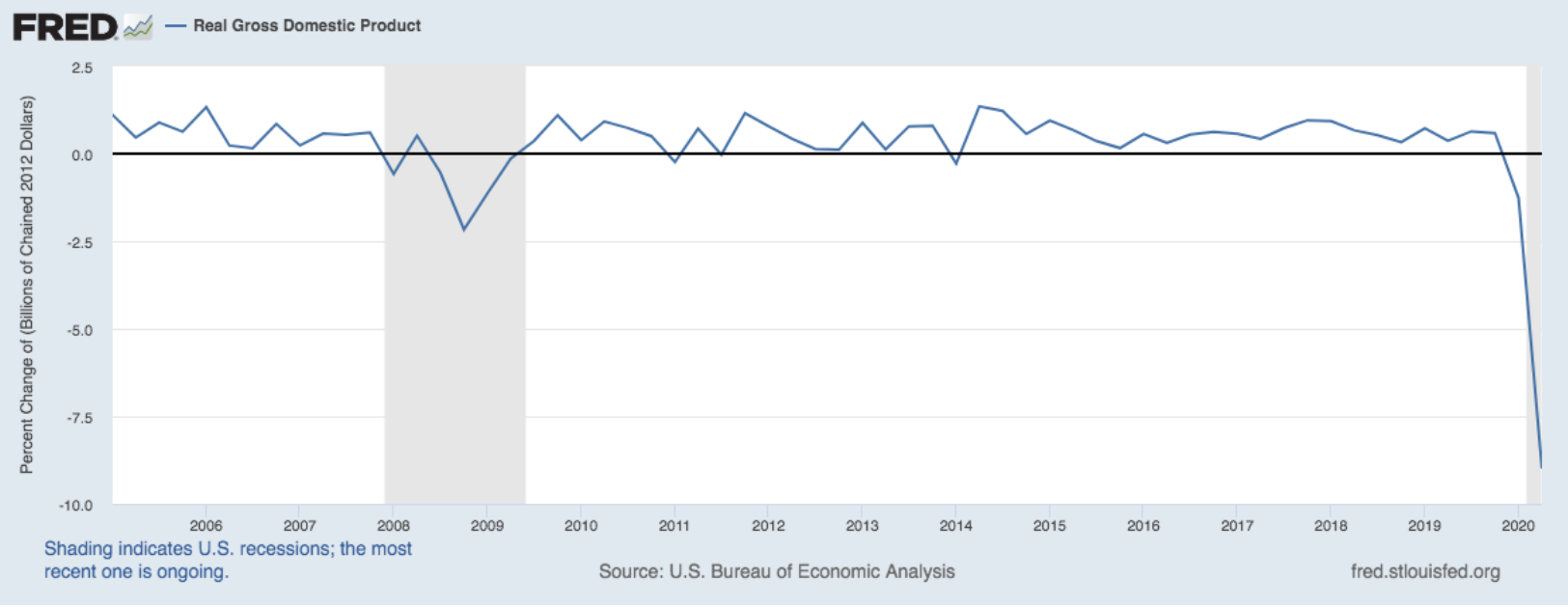}
  \caption{Quarterly data of U.S. real GDP growth rate.}
  \label{fig:GDP}
    \end{subfigure}
    
    \begin{subfigure}{0.8\linewidth}
      \centering
  \includegraphics[width=\linewidth]{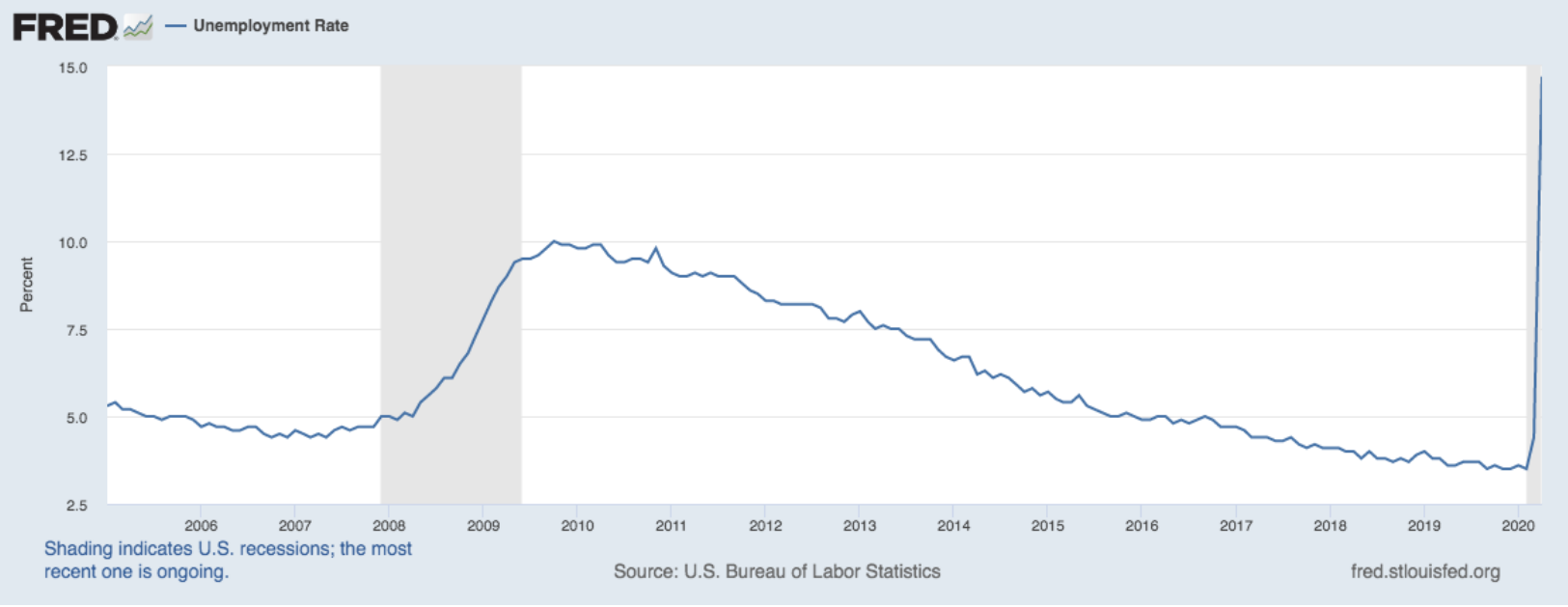}
  \caption{Monthly data of U.S. unemployment rate.}
  \label{fig:unemployment}
    \end{subfigure}
    
        \begin{subfigure}{0.8\linewidth}
      \centering
  \includegraphics[width=\linewidth]{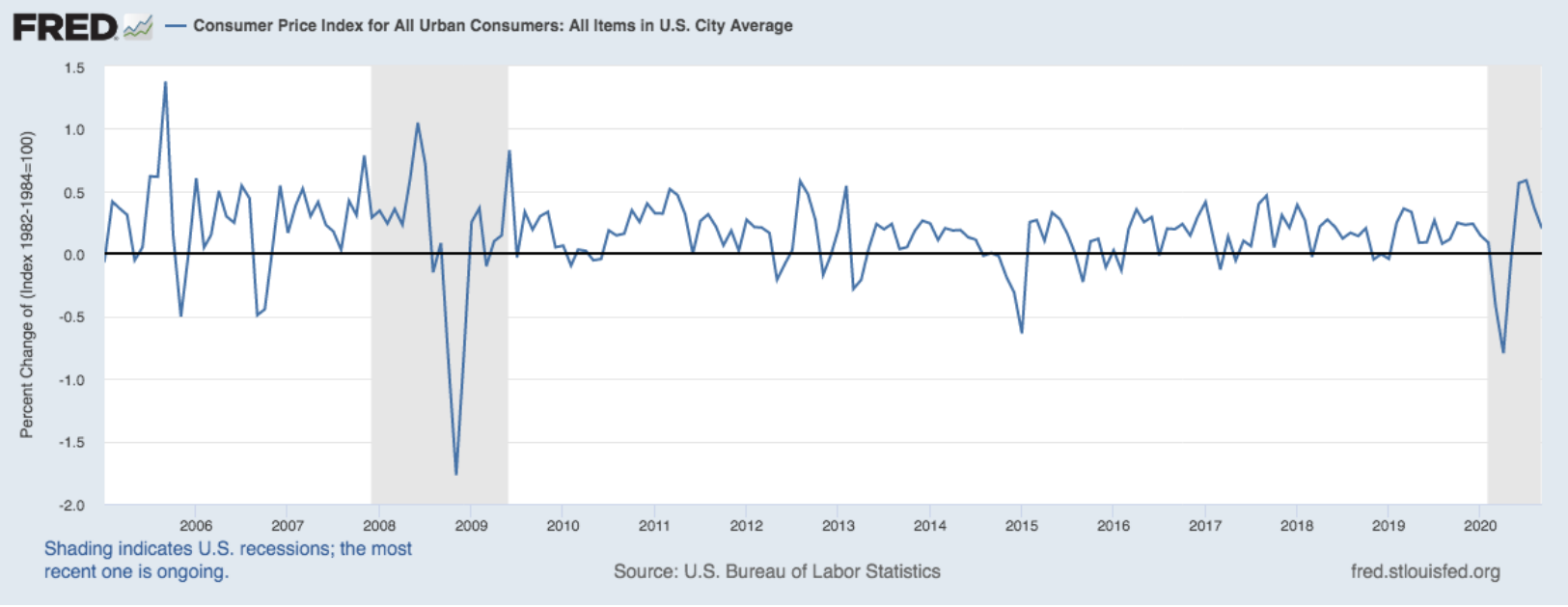}
  \caption{Monthly data of U.S. inflation rate.}
  \label{fig:inflation}
    \end{subfigure}
    
  \caption{Interest rate, real GDP growth rate, unemployment rate, and inflation rate for the period of 2005 and 2020.}
  \label{fig:interest-gdp-unemp}
   \vspace{-0.2in}
\end{figure}

In recent years, the world's most influential central banks include U.S. Federal Reserve, the European Central Bank, the Bank of Japan, and the Bank of England has cut interest rates to near zero in response to the Great Recession and COVID-19 pandemic. Figure \ref{fig:interestrate} shows the monthly data of U.S. interest rate for the period 2005: M1 to 2020: M9 which shows both crises in the shading area. As shown in this Figure, the Federal Reserve cuts interest rate from 4.24 in Dec 2007 to 0.21 in June 2009 (during the Great Recession) and from 1.58 in Feb 2020 to 0.09 in Sep 2020 (during the COVID-19 pandemic).

In fact, central banks including Federal Reserve use Taylor rule \cite{Taylor93} as a measure of monetary policy to adjust interest rate in response to developments in inflation and economic activity. Taylor rule explains that the nominal interest rate ($i_t$) should respond to divergences of actual inflation rates ($\pi_t$) from target inflation rates ($\pi_{t}^{*}$) and of actual GDP($y_t$) from potential GDP ($\overline{y}_{t}$) as below:

 \vspace{-0.15in}
\[
i_t = \pi_t + r_t^* + \alpha_\pi (\pi_t - \pi_t^* ) + \alpha_y (y_t - \overline{y}_t)
\]

where both $\alpha_\pi$ and $\alpha_y$ should be positive (equal to 0.5 in original version of this rule). By using Taylor rule, the Federal Reserve policymakers adopt high interest rate when inflation is above its target (or when real GDP growth rate is higher than its potential level). The Federal Reserve policymakers also adopt low interest rate when inflation is below its target (or when real GDP growth rate is lower than its potential level). Indeed, the Federal Reserve statements cover all discussion about economic indicators and interest rates.

Figure \ref{fig:GDP} shows the quarterly data of U.S. GDP growth rate for the period 2005: Q1 to 2020: Q1 which covers both crises in the shading area. As shown in this figure, the real GDP growth rate in the U.S. was negative in both crises (-2.16 percent in 2008: Q4 and -8.99 in 2020: Q1).  In response to both crises, the Federal Reserve cuts interest rate with the goal of stimulating economic growth.

Figure \ref{fig:unemployment} illustrates the U.S. unemployment rates during the period of 2005: M1 and 2020: M4 which covers the Great Recession and the pandemic. As the figure shows, while the unemployment rate reached close to 9.5\% in June 2010, it has reached to 14.7\% in April 2020, due to the pandemic. Figure \ref{fig:inflation} illustrates monthly data of U.S. inflation rate for the period of 2005: M1-2020: M9. As shown in this figure, the inflation rate sharply had decreased during the Great Recession than the pandemic. During the Great Recession, inflation rate had decreased due to lower demand and lower economic activity. However, the COVID-19 crisis is public health and economic crises both. 

Table \ref{tab:summary-ir-gdp-u-infr} summarizes the extreme values for the indicators shown in Figure \ref{fig:interest-gdp-unemp}, for the periods of the Great Recession and COVID-19. In the case of interest rate, real GDP growth rate, and inflation rate, Table \ref{tab:summary-ir-gdp-u-infr} shows minimum values for the period and the maximum values for unemployment rate.

\begin{table}[h]
    \centering
    \begin{tabular}{|r|r|r|}
    \hline
      &  T{\bf he Great Recession} & {\bf COVID-19} \\
    \hline
            Interest rate &   0.15\% &  0.05\% \\
            Real GDP growth &   -2.16\% &  -8.98\% \\
            Unemployment rate &   9.5\% &   14.7\%\\
            Inflation rate &   -1.77\% &  -0.79\%\\
    \hline
    \end{tabular}

    \caption{Summary of extreme values for indicators.}
    \label{tab:summary-ir-gdp-u-infr}
     \vspace{-0.2in}
\end{table}

\section{Conclusion and Future Work}
\label{sec:conclusion}

In this work, we performed a concern analysis though the application of topic modelling using LDA, with both BOW and tf-idf models. We employed a randomized grid search in order to find a good set of hyperparameters for the number of topics, $\alpha$, and $\eta$ according to the coherence score. Finally, we generated wordclouds to inspect to most important words on each topic, MDS plots for topic embeddings using pyLDAvis to visualize how the topics are configurated in the space, and stackplots to assess the evolution of topic dominance in the statements over time for both models.

Our results show that the topic mixtures obtained using LDA on the statements dataset are responsive to disruptive events such as the Great Recession and COVID-19, for both BOW and tf-idf models, as the topic mixtures during these periods change abruptly; this also agrees with the economic impacts of these events. LDA using BOW shows more dominant topic mixtures, whereas using tf-idf the topics become more balanced in comparison. In terms of interpretability, LDA topics generated using BOW offer a clearer interpretation that can provide useful insights to the economic landscape, whereas LDA topics generated using tf-idf are harder to interpret.

As to the hyperparameter tuning, we found that choosing the set of hyperparameters that maximizes the coherence score might also may lead to a topic configuration that is challenging to interpret. Evidence of this is the negligible dominance of topics retrieved using the BOW model and the overlapping topics generated using the tf-idf model.

As future work, it would be interesting to run the experiments again when more statements are issued, as COVID-19 is still an evolving situation. In addition, other topic modelling techniques could be applied to this dataset.

\section*{Acknowledgment}
This research work is supported by National Science Foundation (NSF) under Grants No: 1723765 and 1821560.



\bibliographystyle{IEEEtran}
\bibliography{IEEEfull,sample-base}

\end{document}